\documentclass[sigconf]{acmart}

\pdfoutput=1
\makeatletter
\def\mdseries@tt{m}
\makeatother 

\usepackage{graphicx}
\usepackage{amsmath, amsfonts, amssymb, amsthm} 
\usepackage{hyperref}           
\usepackage{tikz}
\usetikzlibrary{arrows,petri,topaths,backgrounds,patterns,positioning}
\usepackage{pgfplots}
\usepgfplotslibrary{groupplots}
\usepackage{units}
\usepackage{booktabs}
\usepackage{tabularx}
\usepackage{algorithmic}
\usepackage[linesnumbered,lined,boxed]{algorithm2e}
\usepackage[caption=false]{subfig}
\usepgfplotslibrary{fillbetween}
\newsavebox{\mintedbox}
\usepackage{xcolor}
\usepackage{float}
\usepackage{algorithmic}
\usepackage[linesnumbered,lined,boxed]{algorithm2e}
\usepackage{tabularx} 
\usepackage{booktabs}
\usepackage{multirow}
\usepackage[draft=true]{minted}
\usepackage{mathtools}
\usepackage{setspace}
\usepackage{placeins}
\usepackage{tcolorbox}
\usepackage{xparse}
\tcbuselibrary{minted,breakable,xparse,skins}
\usepackage{listings}
\definecolor{bg}{gray}{0.95}
\usepackage{minted}

\AtBeginDocument{%
	\providecommand\BibTeX{{%
			\normalfont B\kern-0.5em{\scshape i\kern-0.25em b}\kern-0.8em\TeX}}}

\newcommand{\specialcell}[2][c]{%
	\begin{tabular}[#1]{@{}c@{}}#2\end{tabular}}

\copyrightyear{2020}
\acmYear{2020}
\setcopyright{acmlicensed}\acmConference[CIKM '20]{Proceedings of the 29th ACM International Conference on Information and Knowledge Management}{October 19--23, 2020}{Virtual Event, Ireland}
\acmBooktitle{Proceedings of the 29th ACM International Conference on Information and Knowledge Management (CIKM '20), October 19--23, 2020, Virtual Event, Ireland}
\acmPrice{15.00}
\acmDOI{10.1145/3340531.3412758}
\acmISBN{978-1-4503-6859-9/20/10}

\begin{document}

\title{Little Ball of Fur: A Python Library for Graph Sampling}

\author{Benedek Rozemberczki}
\affiliation{%
  \institution{The University of Edinburgh}
  \city{Edinburgh}
  \country{United Kingdom}}
\email{benedek.rozemberczki@ed.ac.uk}

\author{Oliver Kiss}
\affiliation{%
  \institution{Central European University}
  \city{Budapest}
  \country{Hungary}}
\email{kiss\_oliver@phd.ceu.edu}

\author{Rik Sarkar}
\affiliation{%
  \institution{The University of Edinburgh}
  \city{Edinburgh}
  \country{United Kingdom}}
\email{rsarkar@inf.ed.ac.uk}

\begin{abstract}
 Sampling graphs is an important task in data mining. In this paper, we describe \textit{Little Ball of Fur} a Python library that includes more than twenty graph sampling algorithms. Our goal is to make node, edge, and exploration-based network sampling techniques accessible to a large number of professionals, researchers, and students in a single streamlined framework. We created this framework with a focus on a coherent application public interface which has a convenient design, generic input data requirements, and reasonable baseline settings of algorithms. Here we overview these design foundations of the framework in detail with illustrative code snippets. We show the practical usability of the library by estimating various global statistics of social networks and web graphs. Experiments  demonstrate that \textit{Little Ball of Fur} can speed up node and whole graph embedding techniques considerably with mildly deteriorating the predictive value of distilled features.
\end{abstract}

\keywords{network sampling, graph sampling, subsampling, network analytics, graph mining}

\maketitle

\section{Introduction}\label{sec:little_ball_of_fur_introduction}

Modern graph datasets such as social networks and web graphs are large and can be mined to extract detailed insights. However, the large size of the datasets pose fundamental computational challenges on graphs~\cite{large_scale_1,large_scale_2}. 
Exploratory data analysis and computation of basic descriptive statistics can be time consuming on real world graphs. More advanced graph mining techniques such as node and edge classification or clustering can be completely intractable on full size datasets such as web graphs.

One of the fundamental techniques to deal with large datasets is sampling. On simple datasets such as point clouds, sampling preserves most of the distributional features of the data and forms the basis of machine learning~\cite{shalev2014understanding}. However, graphs represent complex interrelations, so that naive sampling can destroy the salient features that constitute the value of the graph data. Graph sampling algorithms therefore need to be sensitive to the various features that are relevant to the downstream tasks. Such features include statistics such as diameter, clustering coefficient~\cite{kleinberg2010networks}, transitivity or degree distribution. In more complex situations, graphs are used for community detection, classification, edge prediction etc~\cite{hamilton2017representation}. A sampling algorithm should be representative with respect to such downstream learning tasks.


Various graph sampling procedures have been proposed with different objectives~\cite{sampling_survey}. The implementation of the graph sampling technique and choice of its parameters used for the subgraph extraction can affect its utility for the task in question. A toolbox of well understood graph sampling techniques can make it easier for researchers and practitioners to easily perform graph sampling, and have consistent reproducible sampling across projects. 
Our goal is to make a large number of graph sampling techniques available to a large audience.

\textbf{Present work.} We release \textit{Little Ball of Fur} -- an open-source Python library for graph sampling. 
This is the first publicly available and documented Python graph sampling library. The general design of our framework is centered around an end-user friendly application public interface which allows for fast paced development and interfacing with other graph mining frameworks.

We achieve this by applying a few core software design principles consistently. Sampling techniques are implemented as individual classes and have pre-parametrized constructors with sensible default settings, which include the number of sampled vertices or edges, a random seed value and hyperparameters of the sampling method itself. Algorithmic details of the sampling procedures are implemented as private methods in order to shield the end-user from the inner mechanics of the algorithm. These concealed workings of samplers rely on the standard Python library and Numpy~\cite{numpy}.  Practically, sampling techniques only provide a single shared public method (sample) which returns the sampled graph. Sampling procedures use NetworkX \cite{networkx} and NetworKit \cite{networkit} graphs as the input and the output adheres to the same widely used generic formats. 

We demonstrate the practical applicability of our framework on various social networks and web graphs (e.g. Facebook, LastFM, Deezer, Wikipedia). We show that our package allows the precise estimation of macro level statistics such as average degree, transitivity, and degree correlation. We provide evidence that the use of sampling routines from \textit{Little Ball of Fur} can reduce the runtime of node and whole graph embedding algorithms. Using these embeddings as input features for node and graph classification tasks we establish that the embeddings learned on the subsampled graphs extract high quality features.

\textbf{Our contributions.} Specifically, the main contributions of our work can be summarized as:

\begin{enumerate}
    \item We present \textit{Little Ball of Fur} a Python graph sampling library which includes various node, edge and exploration based subgraph sampling techniques.
    \item We use code snippets to discuss the design principles which we applied when we developed our package. We examine the presence of standard hyperparameter settings, inner sampling mechanics which are implemented as private methods, the unified application public interface and the use of open-source scientific data structures.
    \item We demonstrate on various real world social network and webgraph datasets how sampling a subgraph with \textit{Little Ball of Fur} affects the estimation accuracy of graph-level macro statistics. We present real world graph mining case studies where using our sampling framework speeds up graph embedding.  
    \item We provide a detailed documentation of our sampling package with a tutorial and case studies with code snippets.
\end{enumerate}

The rest of this paper has the following structure. In Section \ref{sec:little_ball_of_fur_related_work} we overview the relevant literature about graph sampling. This discussion covers node, edge, and exploration sampling techniques, and the possible applications of sampling from graphs. The design principles which we followed when \textit{Little Ball of Fur} was developed are discussed in Section \ref{sec:little_ball_of_fur_design} with samples of illustrative Python code. The subsampling techniques provided by our framework are evaluated in Section \ref{sec:little_ball_of_fur_experiments}. We present results about network statistic estimation performance, machine learning case studies about node and graph classification. The paper concludes with Section \ref{sec:little_ball_of_fur_conclusion} where we discuss our main findings and point out directions for future work. We open sourced the software package and it can be downloaded from \url{https://github.com/benedekrozemberczki/littleballoffur}; the Python package can be installed via the \textit{Python Package Index}. A comprehensive documentation can be accessed at 
\url{https://little-ball-of-fur.readthedocs.io/en/latest/} with a step-by-step tutorial.
\section{Related work}\label{sec:little_ball_of_fur_related_work}
In this section we briefly overview the types of graph subsampling techniques included in \textit{Little Ball of Fur} and the node and graph level representation learning algorithms used for the experimental evaluation of the framework.

\subsection{Graph sampling techniques}
Graph subsampling procedures have three main groups -- node, edge, and exploration-based techniques. We give a brief overview of these techniques in this section.
\subsubsection{Node sampling} Techniques which sample nodes select a set of representative vertices and extract the induced subgraph among the chosen vertices. Nodes can be sampled uniformly without replacement (RN) \cite{randomnode}, proportional to the degree centrality of nodes (RDN) \cite{adamic_degree} or according to the pre-calculated PageRank score of the vertices (PRN) \cite{leskovec2006sampling}. All of these methods assume that the set of vertices and edges in the graph is known ex-ante.
\subsubsection{Edge sampling} The simplest link sampling algorithm retains a randomly selected subset of edges by sampling those uniformly without replacement (RE) while another approach is to randomly select nodes and an edge that belongs to the chosen node (RNE) \cite{randomnodeedge}. These techniques can be hybridized by alternating between node-edge sampling and random edge selection with a parametrized probability (HRNE)  \cite{randomnodeedge}.

By randomly selecting a set of retained edges one implicitly samples nodes. Because of this the random edge selection can be followed up by an induction step (TIES) \cite{inductionsampler} in which the additional edges among chosen nodes are all added. This step can be a partial induction (PIES) \cite{inductionsampler} if the edges were sampled in a streaming fashion and only edges with already sampled nodes are selected in the induction step. 

\subsubsection{Exploration based sampling} Node and edge sampling techniques do not extract representative subsamples of a graph by exploring the neighbourhoods of seed nodes. In comparison exploration based sampling techniques probe the neighborhood of a seed node or a set of seed vertices. 

A group of exploration based sampling techniques uses search strategies on the graph to extract a subsample. The simplest search based strategies include classical traversal methods such as breadth first search (BFS) and depth first search (DFS) \cite{hotplanet}. Snow ball sampling (SB) \cite{snowball} is a restricted version of BFS where a maximum fixed $k$ number of neighbors is traversed. Forest fire (FF) sampling \cite{leskovec2005graphs} is a parametrized stochastic version of SB sampling where the constraint on the maximum number of traversed neighbours only holds in expectation. A local greedy search based technique is community structure expansion sampling \cite{communitystructure} (CSE) which starting with a seeding node adds new nodes to the sampled set which can reach the largest number of unknown nodes. Another simpler search based sampling technique is the random node-neighbor (RNN) \cite{leskovec2006sampling} algorithm which randomly selects a set of seed nodes, takes the neighbors in a single hop and induces the edges of the resulting vertex set. Searching for shortest paths (SP) \cite{shortest_path} between randomly sampled nodes can be used for selecting sequences of nodes and edges to induce a subsampled graph.

A large family of exploration based graph sampling strategies is based on random walks (RW) \cite{gjoka2010walking}. These techniques initiate a random walker on a seed node which traverses the graph and induces a subgraph which is used as the sample. Random walk based sampling has numerous shortcomings and a large number of sampling methods tries to correct for specific limitations.

One of the major limitations is that random walks are inherently biased towards visiting high degree nodes in the graph \cite{sampling_survey}, Metropolis-Hastings random walk (MHRW) \cite{hubler2008metropolis, hastings2} and its rejection constrained variant (RC-MHRW) \cite{rejection} address this bias by making the walker prone to visit lower degree nodes.

Another major shortcoming of random walk based sampling is that the walker might get stuck in the closely knit community of the seed node. There are multiple ways to overcome this. The first one is the use of non-backtracking random walks (NBTRW) \cite{backtracking} which removes the tottering behaviour of random walks. The second one is circulating the neighbors of every node with a vertex specific queue (CNRW) \cite{circulated}. A third strategy involves teleports - the random walker jumps  (RWJ) \cite{ribeiro2010estimating} with a fixed probability to a random node from the current vertex. A fourth approach is to make the walker biased towards weak links by creating a common neighbor aware random walk sampler (CNARW) \cite{commonneighbor} which is biased towards neighbors with low neighborhood overlap. A fifth strategy is using multiple random walkers simultaneously which form a so called frontier of random walkers (FRW) \cite{ribeiro2010estimating}. These techniques can be combined with each other in a modular way to overcome the limitations of random walk based sampling.

There are other possible modifications to traditional random walks which we implemented in \textit{Little Ball of Fur}. One example is random walk with restart (RWR) \cite{leskovec2006sampling}, which is similar to RWJ sampling, but the teleport always ends with the seed node or loop erased random walks (LERW) \cite{wilson} which can sample spanning trees from a source graph uniformly. 
\subsection{Node and whole graph embedding}
Our experimental evaluation includes node and graph classification for which we use features extracted with neighbourhood preserving node embeddings and whole graph embedding techniques. 
\subsubsection{Neighbourhood preserving node embedding} Given a graph $G=(V,E)$ neighbourhood preserving node embedding techniques \cite{deepwalk,line, node2vec, hope, grarep, walklets, diff2vec} learn a function $f: V\to \mathbb{R}^d$ which maps the nodes $ v\in V$ into a $d$ dimensional Euclidean space. In this embedding space a pre-determined notion of node-node proximity is approximately preserved by learning the mapping. The vector representations created by the embedding procedure can be used as input features for node classifiers.  

\subsubsection{Whole graph embedding and statistical fingerprinting} Starting with a set of graphs $\mathcal{G}=(G_1,\dots, G_n)$ whole graph embedding and statistical fingerprinting procedures \cite{graph2vec,gl2vec,vermahunt, netlsd, feather} learn a function $h: \mathcal{G}\to \mathbb{R}^d$ which maps the graphs $G \in \mathcal{G}$ to a $d$ dimensional Euclidean space. In this space those graphs which have similar structural patterns are close to each other. The vector representations distilled by these whole graph embedding techniques are useful inputs for graph classification algorithms. 
\section{Design principles}\label{sec:little_ball_of_fur_design}
We overview the core design principles that we applied when we designed \textit{Little Ball of Fur}. Each design principle is discussed with illustrative examples of Python code which we explain in detail.
\subsection{Encapsulated sampler hyperparameters, random seeding, and parameter inspection}
Graph sampling methods in \textit{Little Ball of Fur} are implemented as individual classes which all inherit from the \textit{Sampler} class. A \textit{Sampler} object is created by using the constructor which has default out-of-the-box hyperparameter settings. These default settings are available in the documentation and can be customized by re-parametrizing the \textit{Sampler} constructor. The hyperparameters of the sampling techniques are stored as \textit{public attributes} of the \textit{Sampler} instance which allows for inspection of the hyperparameter settings by the user. Each graph sampling procedure has a seed parameter -- this value is used to set a random seed for the standard Python and NumPy random number generators. This way the subsample extracted from a specific graph with a fixed seed is always going to be the same.

The code snippet in Figure \ref{fig:inspection_example}
 illustrates the encapsulated hyperparameter and inspection features of the framework. We start the script by importing a simple random walk sampler from the package (line 1). We initialize the first random walk sampler instance without changing the default hyperparameter settings (line 3). As the seed and hyperparameters are exposed we can print the seed parameter which is a public attribute of the sampler (line 4) and we can see the default value of the seed. We create a new instance with a parametrized constructor which sets a new seed (line 6) which modifies the value of the publicly available random seed (line 7).

\begin{figure}[h!]
\begin{minted}[linenos,fontsize=\small,xleftmargin=0.5cm,numbersep=3pt,frame=lines]{python}
from littleballoffur import RandomWalkSampler

sampler = RandomWalkSampler()
print(sampler.seed)

sampler = RandomWalkSampler(seed=41)
print(sampler.seed)
\end{minted}
\caption{Re-parametrizing and initializing the constructor of a random walk sampler by changing the random seed.}\label{fig:inspection_example}
\end{figure}

\subsection{Achieving API consistency and non proliferation of classes}

The graph sampling procedures included in \textit{Little Ball of Fur} are implemented with a consistent application public interface. As we discussed the parametrized constructor is used to create the sampler instance and the samplers all have a single available \textit{public method}. The subsample of the graph is extract by the use of the \textit{sample} method which takes the source graph and calls the private methods of the sampling algorithm.

We limited the number of classes and methods in \textit{Little Ball of Fur} with a straightforward design strategy. First, the graph sampling procedures do not rely on custom data structures to represent the input and output graphs. Second, inheritance from the \textit{Sampler} ensures that private methods which check the input format requirements do not have to be re-implemented on a case-by-case basis for each sampling procedure.
\begin{figure}[h!]
\begin{minted}[linenos,fontsize=\small,xleftmargin=0.5cm,numbersep=3pt,frame=lines]{python}
import networkx as nx
from littleballoffur import RandomWalkSampler

graph = nx.watts_strogatz_graph(1000, 10, 0)  

sampler = RandomWalkSampler()
sampled_graph = sampler.sample(graph)

print(nx.transitivity(sampled_graph))
\end{minted}
\caption{Using a random walk sampler on a Watts-Strogatz graph without changing the default sampler settings.}\label{fig:random_walk_example}
\end{figure}

In Figure \ref{fig:random_walk_example}, first we import \textit{NetworkX} and the random walk sampler from \textit{Little Ball of Fur} (lines 1-2).  Using these libraries we create a Watts-Strogatz graph (line 4) and a random walk sampler with the default hyperparameter settings of the sampling procedure (line 6). We sample a subgraph with the public \textit{sample} method of the random walk sampler (line 7) and print the transitivity calculated for the sampled subgraph (line 8).

\begin{figure}[h!]
\begin{minted}[linenos,fontsize=\small,xleftmargin=0.5cm,numbersep=3pt,frame=lines]{python}
import networkx as nx
from littleballoffur import ForestFireSampler

graph = nx.watts_strogatz_graph(1000, 10, 0)  

sampler = ForestFireSampler()
sampled_graph = sampler.sample(graph)

print(nx.transitivity(sampled_graph))
\end{minted}
\caption{Using a forest fire sampler on a Watts-Strogatz graph without changing the default sampler settings.}\label{fig:forest_fire_example}
\end{figure}

The piece of code presented in Figure \ref{fig:random_walk_example} can be altered seamlessly to perform Forest Fire sampling by modifying the sampler import (line 2) and changing the constructor (line 7) -- these modifications result in the example in Figure \ref{fig:forest_fire_example}.

These illustrative sampling pipelines presented in Figures \ref{fig:random_walk_example} and \ref{fig:forest_fire_example} demonstrate the advantage of maintaining API consistency for the samplers. Changing the graph sampling technique that we used only required minimal modifications to the code. First, we replaced the import of the sampling technique from the \textit{Little Ball of Fur} framework. Second, we used the constructor of the newly chosen sampling technique to create a sampler instance. Finally, we were able to use the shared \textit{sample} method and the same pipeline to calculate the transitivity of the sampled graph. 

\subsection{Backend deployment, standardized dataset ingestion and limitations}
Little Ball of Fur was implemented with a backend wrapper. Sampling procedures can be executed by the \textit{NetworKit} \cite{networkit} or \textit{NetworkX} \cite{networkx} backend libraries depending on the format of the input graph. Basic graph operations such as random neighbor or shortest path retrieval of the backend libraries have standardized naming conventions, data generation and ingestion methods. The generic backed wrapper based design allows for the future inclusion of other general graph manipulation backend libraries such as \textit{GraphTool} \cite{graphtool} or \textit{SNAP} \cite{snap}.  

The shared public \textit{sample} method of the node, edge and exploration based sampling algorithms takes a \textit{NetworkX}/\textit{Networkit} graph as input and the returned subsample is also a \textit{NetworkX}/\textit{Networkit} graph. The subsampling does not change the indexing of the vertices.

The rich ecosystem of graph subsampling methods and the consistent API required that \textit{Little Ball of Fur} was designed with a limited scope and we made restrictive assumptions about the input data used for sampling. Specifically, we assume that vertices are indexed numerically, the first index is 0 and indexing is consecutive. We assume that the graph that is passed to the sample method is undirected and unweighted (edges have unit weights). In addition, we assume that the graph forms a single strongly connected component and orphaned nodes are not present. Heterogeneous, multiplex, multipartite and attributed graphs are not handled by the 1.0 release of the sampling framework.

The sampler classes all inherit private methods that check whether the input graph violates the listed assumptions. These are called within the \textit{sample} method before the sampling process itself starts. When any of the assumptions is violated an error message is displayed for the end-user about the wrong input and the sampling is halted.
\section{Experimental Evaluation}\label{sec:little_ball_of_fur_experiments}
To evaluate the sampling algorithms implemented in \textit{Little Ball of Fur} we perform a number of experiments on real world social networks and webgraphs. Details about these datasets are discussed in Subsection \ref{subsec:datasets}.  We show how randomized spanning tree sampling can be used to speed up node embedding in Subsection \ref{subsec:node_embedding} without reducing predictive performance.  Our ablation study about graph classification in Subsection \ref{subsec:whole_graph_embedding} demonstrates how connected graph subsampling can accelerate the application of whole graph embedding techniques. We present results about estimating graph level descriptive statistics in Subsection \ref{subsec:descriptives}.

\begin{table}[htbp!]

	\centering
	\caption{Statistics of social networks used for comparing sampling and node classification algorithms.}
	\label{tab:little_ball_of_fur_node_statistics}
	
	\centering{\small
		\begin{tabular}{lcccc}
			& \specialcell{\textbf{Facebook}\\\textbf{Page-Page}}  &\specialcell{ \textbf{Wikipedia}\\\textbf{Crocodiles}} & \specialcell{\textbf{LastFM}\\\textbf{Asia}}& \specialcell{\textbf{Deezer}\\\textbf{Hungary}}\\[0.45em]
			\hline
			\textbf{Nodes}          &22,470&11,631&7,624&47,538\\[0.25em]
			\textbf{Density}        &0.0007&0.0025&0.0010&0.0002\\[0.25em]
			\textbf{Transitivity}   &0.2323&0.0261&0.1786&0.0929\\[0.25em]
			\textbf{Diameter}       &15&11&15&12\\[0.25em]
			\textbf{Labels}         &4&2&18&84\\[0.25em]
			\hline
	\end{tabular}}
\end{table}
\vspace{-5mm}

%
		
%

\subsection{Datasets}\label{subsec:datasets}
We use real world social network and webgraph datasets to compare the performance of sampling procedures and test their utility for speeding up classification tasks.
\subsubsection{Node level datasets}
The datasets used for graph statistic estimation and node classification are all available on SNAP \cite{snapnets}, and descriptive statistics can be found in Table \ref{tab:little_ball_of_fur_node_statistics}.
\begin{itemize}
    \item \textbf{Facebook Page-Page} \cite{musae} is a webgraph of verified official Facebook pages. Nodes are pages representing politicians, governmental organizations, television shows and companies while the edges are links between the pages. The related task is multinomial node classification for the 4 page categories.
    \item \textbf{Wikipedia Crocodiles} \cite{musae} is a  webgraph of Wikipedia pages about crocodiles. Nodes are the pages and edges are mutual links between the pages. The potential task is binary node classification.
    \item \textbf{LastFM Asia} \cite{feather} is a social network of LastFM (English music streaming service) users who are located in Asian countries. Nodes are users and links are mutual follower relationships. The task on this dataset is multinomial node classification -- one has to predict the location of the users.
    \item \textbf{Deezer Hungary} \cite{gemsec} is a social network of Hungarian Deezer (French music streaming service) users. Nodes are users located in Hungary and edges are friendships. The relevant task is multi-label multinomial node classification - one has to list the music genres liked by the users. 
\end{itemize}

\subsubsection{Graph level datasets}
Our classification study on subsampled sets of graphs utilized forum threads and small sized social networks of developers. The respective descriptive statistics of these datasets are in Table \ref{fig:little_ball_of_fur_graph_level_statistics}. 

\begin{itemize}
    \item \textbf{Reddit Threads 10K} \cite{karateclub} is a random subsample of 10 thousand graphs from the original Reddit threads datasets. Threads can be discussion and non-discussion based and the task is the binary classification of them according to these two categories.  
    \item \textbf{GitHub StarGazers} \cite{karateclub} is a set of small sized social networks. Each social network is a community of developers who starred a specific machine learning or web development package on Github. The task is to predict the type of the repository based on the community graph. 
\end{itemize}

\vspace{-5mm}
\begin{table}[htbp!]
	\centering
	\caption{Descriptive statistics and size of the graph datasets for graph subsampling and whole graph classification.}
	\label{fig:little_ball_of_fur_graph_level_statistics}
	
	{\footnotesize
\setlength\tabcolsep{4pt} 

\begin{tabular}{cccccccc}
            &        & \multicolumn{2}{c}{\textbf{Nodes}} & \multicolumn{2}{c}{\textbf{Density}} & \multicolumn{2}{c}{\textbf{Diameter}} \\[0.25em]
            \cline{3-8} 
\textbf{Dataset}     & \textbf{Graphs} & \textbf{Min}         & \textbf{Max}         & \textbf{Min}          & \textbf{Max}          & \textbf{Min}           & \textbf{Max}          \\[0.25em]\hline
\textbf{Reddit Threads 10K} &   10,000     &       11      &         97    & 0.021             &  0.291     & 2               &   22           \\[0.7em]
\textbf{GitHub StarGazers} & 12,725      &    10        &957      & 0.003     &       0.561            &     2          &18              \\[0.25em]
			\hline
		\end{tabular}}
		
	\end{table}

\vspace{-5mm}

\subsection{Node classification with randomly sampled spanning tree embeddings of networks}\label{subsec:node_embedding}
Node embedding vectors \cite{deepwalk,walklets} are useful compact descriptors of vertex neighbourhoods when it comes to solving classification problems. In traditional classification scenarios the whole graph is used to learn the node embedding vectors. In this experiment we study a situation where the embedding vectors are learned from a randomly sampled spanning tree of the original graph. We compare the predictive value of node embeddings learned on the whole graph with ones learned from spanning trees extracted with randomized BFS \cite{randomnodeedge}, DFS \cite{randomnodeedge} and LERW \cite{wilson}. The main advantage of randomized spanning trees is that storing the whole graph requires $\mathcal{O}(|E|)$ memory when we learn the node embedding. In contrast storing a sampled spanning tree only requires $\mathcal{O}(|V|)$ space.

\subsubsection{Experimental settings} The experimental pipeline which we used for node classification has four stages. 
\begin{enumerate}

\item \textit{Graph sampling.} The BFS, DFS and LERW sample based embeddings start with the extraction of a random spanning tree using \textit{Little Ball of Fur.} This sample is fed to the embedding procedure.
\item \textit{Upstream model.} The sampled graph is fed to the unsupervised upstream models DeepWalk \cite{deepwalk} and Walklets \cite{walklets} which learn the neighbourhood preserving node embedding vectors. We used the Karate Club \cite{karateclub} implementation of these models with the default hyperparameter settings. 
\item \textit{Downstream model.} We inputted the node embedding vectors as input features for a logistic regression classifier -- we used the scikit-learn implementation \cite{scikit} with the default hyperparameter settings. The downstream models were trained with a varying ratio of nodes.
\item \textit{Evaluation.} We report average AUC values on the test set calculated from a 100 seeded sampling, embedding and downstream model training runs. 
\end{enumerate}
\begin{figure}[ht!]
\centering
\begin{tikzpicture}[scale=0.24,transform shape]
\tikzset{font={\fontsize{22pt}{12}\selectfont}}
\begin{groupplot}[group style={group size=2 by 1,
		          horizontal sep=110pt,
		          vertical sep=70pt,ylabels at=edge left},
	              width=0.9\textwidth,
	              height=0.65\textwidth,
	              grid=major,
	              grid style={dashed, gray!40},
	              scaled ticks=false,
	              inner axis line style={-stealth}]
	\nextgroupplot[ytick={0.5,0.6,0.7,0.8,0.9,1.0},
	xtick={1,3,5,7,9},
	ymin=0.5,ymax=1.0,
	xlabel=\% of nodes used for training,
	ylabel=Area under the curve,
	ymin=0.5,
	ymax=1.0,
	enlargelimits=0.1,
	ylabel style={
    yshift=4ex,},
	xlabel style={
    yshift=-3ex,},
	legend style = { column sep = 10pt, legend columns = -1, legend to name = grouplegend, title = DeepWalk}]
	
	\addplot[mark=triangle*,opacity=0.8,mark options={black,fill=red},mark size=7pt]
	coordinates {
(1,0.7409)
(2,0.7757)
(3,0.7995)
(4,0.8216)
(5,0.8393)
(6,0.8444)
(7,0.8452)
(8,0.8539)
(9,0.8548)
	};\addlegendentry{Full Graph}%
	\addplot[mark=diamond*,opacity=0.8,mark options={black,fill=blue},mark size=7pt]
	coordinates {
(1,0.5471)
(2,0.5761)
(3,0.5884)
(4,0.6124)
(5,0.6049)
(6,0.6053)
(7,0.635)
(8,0.6444)
(9,0.644)

	};\addlegendentry{Breadth First Search}%
	\addplot[mark=*,opacity=0.8,mark options={black,fill=green},mark size=5pt]
	coordinates {
(1,0.5098)
(2,0.517)
(3,0.5229)
(4,0.5296)
(5,0.5243)
(6,0.5269)
(7,0.5359)
(8,0.5337)
(9,0.5349)

	};\addlegendentry{Depth First Search}%
	
	\addplot[mark=square*,opacity=0.8,mark options={black,fill=yellow},mark size=5pt]
	coordinates {
(1,0.5205)
(2,0.5315)
(3,0.5327)
(4,0.5369)
(5,0.5479)
(6,0.5596)
(7,0.5564)
(8,0.5659)
(9,0.5602)

	};\addlegendentry{Loop Erased Random Walk}%

	\nextgroupplot[ytick={0.5,0.6,0.7,0.8,0.9,1.0},
	xtick={1,3,5,7,9},
	ymin=0.5,ymax=1.0,
	xlabel=\% of nodes used for training,
	ylabel=Area under the curve,
	enlargelimits=0.1,
	ylabel style={
    yshift=4ex,},
	xlabel style={
    yshift=-3ex,},
	legend style = { column sep = 10pt, legend columns = -1, legend to name = grouplegend, title = Walklets}]
	
	\addplot[mark=triangle*,opacity=0.8,mark options={black,fill=red},mark size=7pt]
	coordinates {
(1,0.9332)
(2,0.9477)
(3,0.9552)
(4,0.9601)
(5,0.9619)
(6,0.9619)
(7,0.9645)
(8,0.9663)
(9,0.9671)
	};\addlegendentry{Full Graph}%
	\addplot[mark=diamond*,opacity=0.8,mark options={black,fill=blue},mark size=7pt]
	coordinates {
(1,0.6463)
(2,0.6721)
(3,0.6631)
(4,0.704)
(5,0.7093)
(6,0.706)
(7,0.7444)
(8,0.749)
(9,0.7347)

	};\addlegendentry{Breadth First Search}%
	\addplot[mark=*,opacity=0.8,mark options={black,fill=green},mark size=5pt]
	coordinates {
(1,0.5234)
(2,0.5308)
(3,0.5325)
(4,0.5474)
(5,0.5422)
(6,0.5506)
(7,0.5601)
(8,0.5574)
(9,0.5583)
	};\addlegendentry{Depth First Search}%
	
	\addplot[mark=square*,opacity=0.8,mark options={black,fill=yellow},mark size=5pt]
	coordinates {
(1,0.5489)
(2,0.5653)
(3,0.5836)
(4,0.592)
(5,0.611)
(6,0.6125)
(7,0.6168)
(8,0.6226)
(9,0.6076)
	};\addlegendentry{Loop Erased Random Walk}%

	\end{groupplot}	
	\node at ($(group c1r1) + (9.0cm,-8.0cm)$) {\ref{grouplegend}}; 
	\end{tikzpicture}
\caption{Node classification performance on the Facebook Page-Page graph \cite{musae} evaluated by average AUC scores on the test set calculated from a 100 seeded experimental runs.  }\label{fig:facebook}
\end{figure}
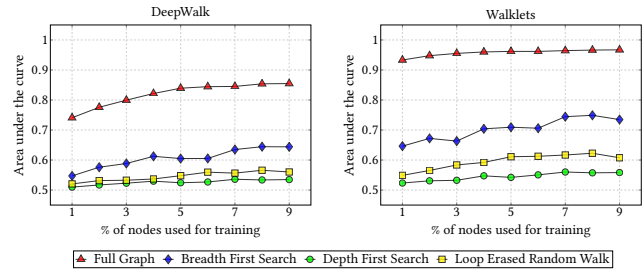
\vspace{-5mm}

\subsubsection{Experimental results} 
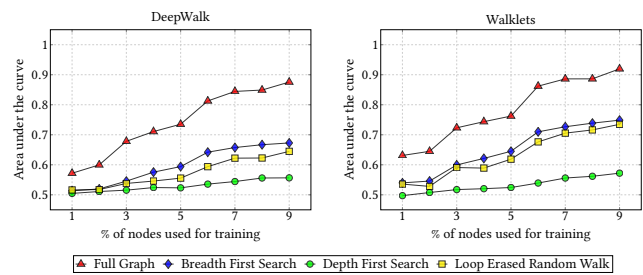
\begin{figure}[ht!]
\centering
\begin{tikzpicture}[scale=0.24,transform shape]
\tikzset{font={\fontsize{22pt}{12}\selectfont}}
\begin{groupplot}[group style={group size=2 by 1,
		          horizontal sep=110pt,
		          vertical sep=70pt,ylabels at=edge left},
	              width=0.9\textwidth,
	              height=0.65\textwidth,
	              grid=major,
	              grid style={dashed, gray!40},
	              scaled ticks=false,
	              inner axis line style={-stealth}]
	\nextgroupplot[ytick={0.5,0.6,0.7,0.8,0.9,1.0},
	xtick={1,3,5,7,9},
	ymin=0.5,ymax=1.0,
	xlabel=\% of nodes used for training,
	ylabel=Area under the curve,
	enlargelimits=0.1,
	ylabel style={
    yshift=4ex,},
	xlabel style={
    yshift=-3ex,},
	legend style = { column sep = 10pt, legend columns = -1, legend to name = grouplegend, title = DeepWalk}]
	
	\addplot[mark=triangle*,opacity=0.8,mark options={black,fill=red},mark size=7pt]
	coordinates {
(1,0.5719)
(2,0.5997)
(3,0.6782)
(4,0.7107)
(5,0.7352)
(6,0.8127)
(7,0.8444)
(8,0.849)
(9,0.8756)

	};\addlegendentry{Full Graph}%
	\addplot[mark=diamond*,opacity=0.8,mark options={black,fill=blue},mark size=7pt]
	coordinates {
(1,0.5133)
(2,0.5196)
(3,0.5455)
(4,0.5755)
(5,0.594)
(6,0.642)
(7,0.6575)
(8,0.6669)
(9,0.6728)

	};\addlegendentry{Breadth First Search}%
	\addplot[mark=*,opacity=0.8,mark options={black,fill=green},mark size=5pt]
	coordinates {
(1,0.5047)
(2,0.511)
(3,0.5155)
(4,0.5242)
(5,0.5232)
(6,0.5356)
(7,0.5442)
(8,0.556)
(9,0.5565)

	};\addlegendentry{Depth First Search}%
	
	\addplot[mark=square*,opacity=0.8,mark options={black,fill=yellow},mark size=5pt]
	coordinates {
(1,0.5162)
(2,0.5175)
(3,0.5379)
(4,0.5458)
(5,0.5556)
(6,0.5938)
(7,0.6221)
(8,0.6226)
(9,0.6447)

	};\addlegendentry{Loop Erased Random Walk}%

	\nextgroupplot[ytick={0.5,0.6,0.7,0.8,0.9,1.0},
	xtick={1,3,5,7,9},
	ymin=0.5,ymax=1.0,
	xlabel=\% of nodes used for training,
	ylabel=Area under the curve,
	enlargelimits=0.1,
	ylabel style={
    yshift=4ex,},
	xlabel style={
    yshift=-3ex,},
	legend style = { column sep = 10pt, legend columns = -1, legend to name = grouplegend, title = Walklets}]
	
	\addplot[mark=triangle*,opacity=0.8,mark options={black,fill=red},mark size=7pt]
	coordinates {
(1,0.6311)
(2,0.6449)
(3,0.7232)
(4,0.7436)
(5,0.7623)
(6,0.862)
(7,0.8862)
(8,0.8863)
(9,0.9196)
	};\addlegendentry{Full Graph}%
	\addplot[mark=diamond*,opacity=0.8,mark options={black,fill=blue},mark size=7pt]
	coordinates {
(1,0.5398)
(2,0.5458)
(3,0.5995)
(4,0.6212)
(5,0.645)
(6,0.7098)
(7,0.7267)
(8,0.7387)
(9,0.7487)
	};\addlegendentry{Breadth First Search}%
	\addplot[mark=*,opacity=0.8,mark options={black,fill=green},mark size=5pt]
	coordinates {
(1,0.4972)
(2,0.5076)
(3,0.5174)
(4,0.5204)
(5,0.5243)
(6,0.5392)
(7,0.5559)
(8,0.5617)
(9,0.5721)

	};\addlegendentry{Depth First Search}%
	
	\addplot[mark=square*,opacity=0.8,mark options={black,fill=yellow},mark size=5pt]
	coordinates {
(1,0.5357)
(2,0.528)
(3,0.5908)
(4,0.589)
(5,0.6187)
(6,0.6763)
(7,0.7054)
(8,0.7162)
(9,0.7348)
	};\addlegendentry{Loop Erased Random Walk}%

	\end{groupplot}	
	\node at ($(group c1r1) + (9.0cm,-8.0cm)$) {\ref{grouplegend}}; 
	\end{tikzpicture}
\caption{Node classification performance on the LastFM Asia graph \cite{feather} evaluated by average AUC scores on the test set calculated from a 100 seeded experimental runs.  }\label{fig:lastfm}
\end{figure}

We report the predictive performance for the Facebook Page-Page and LastFM Asia graphs respectively on Figures \ref{fig:facebook} and \ref{fig:lastfm}. First, we see that the features extracted from the BFS, DFS and LERW sampled spanning trees are less valuable for node classification based on the predictive performance on these two social networks. In plain words node embeddings of randomly sampled spanning trees produce inferior features. Second, the marginal gains of additional training data are smaller when the embedding is learned from a subsampled graph. Third, DFS sampled node embedding features have a considerably lower quality compared to the BFS and LERW sampled node embedding features. Finally, the Walklets based higher order proximity preserving embeddings have a superior predictive performance compared to the DeepWalk based ones even when the graph being embedded is a randomly sampled spanning tree of the source graph.

\vspace{-5mm}

\subsection{An ablation study of graph classification}\label{subsec:whole_graph_embedding}

Graph classification procedures use the whole graph embedding vectors as input to discriminate between graphs based on structural patterns. Using subsamples of the graphs and extracting structural patterns from those can speed up this classification process. We will investigate how exploration based sampling techniques perform when they are used to obtain the samples used for the embedding.

\input{plots_and_tables/graph_classification_auc}

\subsubsection{Experimental settings.} The data processing which we used for the evaluation of graph classification performance has four stages. 
\begin{enumerate}

\item \textit{Graph sampling.} We sample both datasets using the RW \cite{gjoka2010walking} and RWR \cite{leskovec2006sampling} methods implemented in \textit{Little Ball of Fur} 100 times for each retainment rate with different random seeds. Using these algorithms ensures that the graphs' connectivity patterns are unchanged. 
\item \textit{Upstream model.} Following the sampling, all of the samples are embedded using the Graph2Vec \cite{graph2vec} algorithm. This procedure uses the presence of subtrees as structural patterns.
\item \textit{Downstream model.} With the embedding vectors as covariates, we estimate a logistic regression for each dataset and retainment rate. We rely on the scikit-learn implementation \cite{scikit} of the classifier with the default hyperparameter settings. We use 80\% of the graphs to train the classifier.
\item \textit{Evaluation.} The classification performance is evaluated using the AUC based on the remaining 20\% of graphs which form the test set.
\end{enumerate}

\input{plots_and_tables/graph_classification_runtime}
\subsubsection{Experimental results.} We report mean AUC values along with a standard deviation band in Figure \ref{fig:auc} for the Reddit Threads and Github Stargazers datasets with the Random Walk and Random Walk with Restart sampling methods. Lower retainment rate is associated with a lower classification performance, as it can be expected. The more ragged, step function-like pattern observed in case of the Reddit threads dataset is likely to be due to the interplay of structural pattern downsampling and the generally smaller graphs in the dataset.

We report the mean runtime of the graph embedding process with a band of standard deviations in Figure \ref{fig:runtime}. As we decrease the retainment rate, a significant decrease in runtime is prevalent. There is a clear trade-off between runtime and predictive performance. It is, however, worth noting that while the runtime associated with the 50\% retainment rate is, in most cases close to half of the runtime using the whole graphs, the loss in classification power in case of the Reddit Threads dataset is less significant. 

\subsection{Estimating descriptive statistics}\label{subsec:descriptives}
A traditional task for the evaluation of graph sampling algorithms is the estimation of graph level descriptive statistics. The graph level descriptive statistic is calculated for the sampled graph and it is compared to the ground truth value which is calculated based on the whole set of nodes and edges. A well performing sampling procedure is ought to give a precise estimate for the graph level quantity of interest with a reasonably small subsample of the graph.

\subsubsection{Experimental settings} The pipeline used for estimating the graph level descriptive statistics had two stages.
\begin{enumerate}
\item \textit{Graph sampling.} Node and exploration based sampling procedures sample 50\% of vertices, while the edge sampling techniques select 50\% of the edges to extract a subgraph.
\item \textit{Descriptive statistic estimation.} We calculated the average of the clustering coefficient (transitivity), average node degree and the degree correlation for the sampled graphs. We did 10 seeded experimental runs to get an estimate of the statistics and calculated the standard error around these averages.
\end{enumerate}
\subsubsection{Experimental results} The ground truth and estimated descriptive statistics are enclosed in Table \ref{tab:estimates} for all of the node level datasets. Blocks of rows correspond to node, edge, random walk based and non random walk based exploration sampling algorithms. In each block of methods bold numbers denote the best performing sampling technique (closest to the ground truth) in a given category for a specific estimated descriptive statistic and dataset. 

We can make a few generic observations about the quality of descriptive statistic estimates. First, there is not a clearly superior sampling technique. This holds generally and within all of the main categories of considered algorithms. Specifically, there is not a superior node, edge or expansion based sampling procedure. Second, the induction based edge sampling techniques (TIES and PIES) give a good estimate of the statistics, but the induction step includes more than 50\% of the edges. Because of this, the obtained good estimation performance is somewhat misleading as the majority of edges is still retained after the induction step. Third, edge sampling algorithms sometimes fail to estimate the direction of the degree correlation properly. Finally, the random walk based techniques generally tend to overestimate the average degree. This is not surprising considering that these are biased towards high degree nodes. 

\begin{table*}[h!]
\caption{Ground truth and estimated descriptive statistics of the web graphs and social networks. We calculated average statistics from 10 seeded experimental runs and included the standard errors below the mean. We included the ground truth values based on the whole graph (first block) with estimates obtained with node (second block), edge (third block) and exploration (fourth and fifth blocks) sampling algorithms. Bold numbers denote for each category the best estimate for a given dataset.}\label{tab:estimates}
{\footnotesize
\setlength{\tabcolsep}{0.4em}
{\centering 
\begin{tabular}{l ccc ccc ccc ccc}
             & \multicolumn{3}{c}{\textbf{Facebook Page-Page}}  & \multicolumn{3}{c}{\textbf{Wikipedia Crocodiles}} & \multicolumn{3}{c}{\textbf{LastFM Asia}} & \multicolumn{3}{c}{\textbf{Deezer Hungary}} \\ \cmidrule(lr){2-4}
             \cmidrule(lr){5-7}
             \cmidrule(lr){8-10}
             \cmidrule(lr){11-13}
             & \specialcell{\textbf{Clustering}\\ \textbf{Coefficient}}    & \specialcell{\textbf{Average}\\ \textbf{Degree}}    & \specialcell{\textbf{Degree} \\\textbf{Correlation}} & \specialcell{\textbf{Clustering}\\ \textbf{Coefficient}}    & \specialcell{\textbf{Average} \\\textbf{Degree}}   &\specialcell{\textbf{Degree} \\\textbf{Correlation}}  & \specialcell{\textbf{Clustering}\\ \textbf{Coefficient}}       & \specialcell{\textbf{Average} \\\textbf{Degree}}   &\specialcell{\textbf{Degree} \\\textbf{Correlation}} & \specialcell{\textbf{Clustering}\\ \textbf{Coefficient}}   & \specialcell{\textbf{Average} \\\textbf{Degree}}   & \specialcell{\textbf{Degree} \\\textbf{Correlation}} \\ \cmidrule(lr){2-13}
\textbf{Truth} & $0.232$  &    $15.205$  &  $0.085$ & $0.026$  &     $29.365$     &  $-0.277$        &   $0.179$        &  $7.294$        &     $0.017$     &     $0.093$     &     $9.377$     &    $0.207$     \\[0.35em] \hline
\textbf{RN} \cite{randomnode} &  $\underset{0.002}{\textbf{0.229}}$      &   $\underset{0.060}{7.531}$        &   $\underset{0.003}{\textbf{0.070}}$       &  $\underset{0.001}{\textbf{0.028}}$        &      $\underset{0.388}{14.293}$    &   $\underset{\,\,\,0.008}{\textbf{-0.284}}$       &  $\underset{0.004}{\textbf{0.177}}$         &  $\underset{0.032}{3.642}$        &     $\underset{0.010}{\textbf{0.020}}$     &    $\underset{0.001}{\textbf{0.092}}$      &  $\underset{0.010}{4.699}$        &    $\underset{0.002}{0.190}$     \\[0.55em]
\textbf{DRN} \cite{adamic_degree}  &    $\underset{0.001}{0.261}$       &  $\underset{0.021}{21.514}$&$\underset{0.001}{0.080}$&  $\underset{0.001}{0.038}$        &  $\underset{0.054}{40.750}$        &  $\underset{\,\,\,0.001}{-0.324}$        &   $\underset{0.001}{0.231}$        &         $\underset{0.020}{9.531}$ &     $\underset{0.001}{0.045}$     &        $\underset{0.001}{0.102}$  &    $\underset{0.007}{\textbf{8.551}}$      &     $\underset{0.001}{\textbf{0.211}}$    \\[0.55em]
\textbf{PRN} \cite{leskovec2006sampling}  &     $\underset{0.001}{0.270}$      & $\underset{0.032}{\textbf{16.228}}$          &   $\underset{0.001}{0.136}$       &     $\underset{0.001}{0.049}$     &  $\underset{0.074}{\textbf{32.370}}$        &  $\underset{\,\,\,0.001}{-0.290}$        &  $\underset{0.001}{0.236}$         &     $\underset{0.022}{\textbf{8.209}}$     &    $\underset{0.002}{0.064}$      &   $\underset{0.001}{0.098}$       &     $\underset{0.007}{7.251}$     &   $\underset{0.001}{0.231}$     \\[0.55em]
\hline
\textbf{RE}  \cite{randomnodeedge}  &$\underset{0.001}{0.116}$&$\underset{0.004}{8.470}$&$\underset{0.001}{\textbf{0.084}}$&$\underset{0.001}{0.013}$&$\underset{0.009}{15.462}$&$\underset{\,\,\,0.001}{\textbf{-0.277}}$&$\underset{0.001}{0.090}$&$\underset{0.009}{4.422}$&$\underset{0.002}{\textbf{0.019}}$&$\underset{0.001}{0.046}$&$\underset{0.001}{5.018}$&$\underset{0.001}{0.183}$\\[0.55em]
\textbf{RNE} \cite{randomnodeedge}  & $\underset{0.001}{0.092}$&$\underset{0.001}{7.602}$&$\underset{0.001}{-0.075}$&$\underset{0.001}{0.007}$&$\underset{0.001}{14.682}$&$\underset{\,\,\,0.001}{-0.231}$&$\underset{0.001}{0.046}$&$\underset{0.001}{3.674}$&$\underset{0.001}{-0.108}$&$\underset{0.001}{0.042}$&$\underset{0.001}{4.701}$&$\underset{0.001}{0.056}$       \\[0.55em]
\textbf{HRNE} \cite{randomnodeedge} &$\underset{0.001}{0.081}$&$\underset{0.002}{7.194}$&$\underset{0.001}{-0.005}$&$\underset{0.001}{0.007}$&$\underset{0.005}{13.550}$&
$\underset{\,\,\,0.001}{-0.234}$&$\underset{0.001}{0.046}$&$\underset{0.003}{3.562}$&$\underset{0.002}{-0.070}$&$\underset{0.001}{0.039}$&$\underset{0.001}{4.501}$&$\underset{0.001}{0.095}$    \\[0.55em]
\textbf{TIES} \cite{inductionsampler}  & $\underset{0.001}{0.235}$&$\underset{0.007}{16.564}$&$\underset{0.001}{0.083}$ & $\underset{0.001}{\textbf{0.026}}$&$\underset{\,\,\,0.016}{30.720}$&$\underset{0.001}{-0.278}$
 &   $\underset{0.001}{0.190}$&$\underset{0.010}{8.218}$&$\underset{0.001}{0.027}$   &$\underset{0.001}{\textbf{0.094}}$&$\underset{0.001}{\textbf{9.748}}$&$\underset{0.001}{0.204}$      \\[0.55em]
\textbf{PIES} \cite{inductionsampler} & $\underset{0.001}{\textbf{0.231}}$&$\underset{0.008}{\textbf{15.357}}$&$\underset{0.001}{0.087}$&$\underset{0.001}{\textbf{0.026}}$&$\underset{0.015}{\textbf{29.142}}$&$\underset{\,\,\,0.001}{-0.283}$&$\underset{0.001}{\textbf{0.186}}$&$\underset{0.010}{\textbf{7.247}}$&$\underset{0.001}{0.037}$&$\underset{0.001}{0.086}$&$\underset{0.003}{8.501}$&$\underset{0.001}{\textbf{0.209}}$   \\[0.45em]\hline

\textbf{RW} \cite{gjoka2010walking}  &    $\underset{0.001}{0.255}$       &    $\underset{0.022}{22.535}$       &   $\underset{0.001}{0.073}$       &$\underset{0.001}{0.036}$&$\underset{0.196}{41.648}$&$\underset{\,\,\,0.001}{-0.325}$&$\underset{0.001}{0.224}$&$\underset{0.021}{9.878}$&$\underset{0.003}{0.039}$&$\underset{0.001}{0.104}$&$\underset{0.008}{9.221}$&$\underset{0.001}{0.218}$     \\[0.55em]
\textbf{RWR} \cite{leskovec2006sampling} &$\underset{0.003}{0.253}$&$\underset{0.293}{20.282}$&$\underset{0.007}{0.092}$& $\underset{0.003}{0.043}$&$\underset{0.549}{38.967}$&$\underset{\,\,\,0.006}{-0.313}$&$\underset{0.003}{0.222}$&$\underset{0.087}{9.078}$&$\underset{0.004}{0.036}$&$\underset{0.001}{0.098}$&$\underset{0.082}{9.122}$&$\underset{0.003}{\textbf{0.213}}$    \\[0.55em]
\textbf{RWJ}  \cite{ribeiro2010estimating} & $\underset{0.001}{0.271}$&$\underset{0.036}{18.615}$&$\underset{0.001}{0.123}$&$\underset{0.001}{0.047}$&$\underset{0.074}{34.475}$&$\underset{\,\,\,0.001}{-0.297}$&$\underset{0.001}{0.233}$&$\underset{0.032}{9.012}$&$\underset{0.003}{0.067}$& $\underset{0.001}{0.102}$&$\underset{0.016}{8.351}$&$\underset{0.002}{0.233}$    \\[0.55em]
\textbf{MHRW} \cite{hubler2008metropolis, hastings2}  & $\underset{0.002}{0.280}$&$\underset{0.113}{\textbf{17.903}}$&$\underset{0.003}{0.134}$&$\underset{0.002}{0.119}$&$\underset{0.223}{\textbf{29.914}}$&$\underset{\,\,\,0.004}{-0.146}$&$\underset{0.001}{0.232}$&$\underset{0.041}{\textbf{8.854}}$&$\underset{0.007}{0.102}$&$\underset{0.001}{0.106}$&$\underset{0.023}{7.761}$&$\underset{0.003}{0.242}$      \\[0.55em]
\textbf{RC-MHRW} \cite{rejection}  & $\underset{0.001}{0.266}$&$\underset{0.064}{21.070}$&$\underset{0.002}{0.106}$&$\underset{0.001}{0.072}$&$\underset{0.159}{35.758}$&$\underset{\,\,\,0.002}{-0.254}$&$\underset{0.001}{0.232}$&$\underset{0.031}{9.594}$&$\underset{0.003}{0.078}$&$\underset{0.001}{0.103}$&$\underset{0.021}{\textbf{8.553}}$&$\underset{0.002}{0.237}$      \\[0.55em]
\textbf{FRW}  \cite{ribeiro2010estimating} & $\underset{0.001}{0.063}$&$\underset{0.059}{5.745}$&$\underset{0.004}{0.069}$&$\underset{0.001}{0.004}$&$\underset{0.058}{5.813}$&$\underset{\,\,\,0.003}{\textbf{-0.280}}$&$\underset{0.001}{0.084}$&$\underset{0.032}{4.485}$&$\underset{0.008}{\textbf{0.018}}$&$\underset{0.001}{0.033}$&$\underset{0.005}{3.243}$&$\underset{0.002}{0.091}$      \\[0.55em]
\textbf{CNRW} \cite{circulated}  & $\underset{0.001}{0.255}$&$\underset{0.038}{22.590}$&$\underset{0.001}{0.072}$&$\underset{0.001}{0.037}$&$\underset{0.104}{41.645}$&$\underset{\,\,\,0.001}{-0.324}$&$\underset{0.001}{0.223}$&$\underset{0.018}{9.924}$&$\underset{0.002}{0.036}$&$\underset{0.001}{0.104}$&$\underset{0.017}{9.254}$&$\underset{0.001}{0.218}$       \\[0.55em]
\textbf{CNARW} \cite{commonneighbor}  & $\underset{0.001}{\textbf{0.239}}$&$\underset{0.033}{21.117}$&$\underset{0.001}{\textbf{0.082}}$&$\underset{0.001}{\textbf{0.026}}$&$\underset{0.101}{41.064}$&$\underset{\,\,\,0.001}{-0.348}$&$\underset{0.001}{\textbf{0.220}}$&$\underset{0.032}{9.508}$&$\underset{0.002}{0.052}$&$\underset{0.001}{\textbf{0.094}}$&$\underset{0.013}{9.140}$&$\underset{0.001}{0.218}$  \\[0.55em]
\textbf{NBT-RW} \cite{backtracking}  & $\underset{0.001}{0.257}$&$\underset{0.048}{22.353}$&$\underset{0.001}{0.076}$&$\underset{0.001}{0.038}$&$\underset{0.144}{41.264}$&
$\underset{\,\,\,0.001}{-0.322}$&$\underset{0.001}{0.226}$&$\underset{0.027}{9.818}$&$\underset{0.002}{0.049}$&$\underset{0.001}{0.104}$&$\underset{0.010}{9.106}$&$\underset{0.001}{0.230}$  \\[0.55em]
\hline
\textbf{SB}  \cite{snowball} &$\underset{0.002}{\textbf{0.238}}$&$\underset{0.223}{20.671}$&$\underset{0.004}{0.069}$&$\underset{0.004}{0.057}$&$\underset{0.576}{37.278}$&$\underset{\,\,\,0.009}{-0.292}$&$\underset{0.002}{0.207}$&$\underset{0.121}{9.131}$&$\underset{0.005}{-0.008}$&$\underset{0.001}{\textbf{0.093}}$&$\underset{0.103}{\textbf{9.913}}$&$\underset{0.003}{0.122}$  \\[0.55em]
\textbf{FF} \cite{leskovec2005graphs}  &  $\underset{0.001}{\textbf{0.238}}$         & $\underset{0.089}{19.219}$          &  $\underset{0.002}{\textbf{0.079}}$        & $\underset{0.002}{0.074}$         &   $\underset{0.262}{\textbf{33.190}}$       &  $\underset{\,\,\,0.006}{-0.227}$        &     $\underset{0.001}{0.204}$      & $\underset{0.025}{9.034}$         &  $\underset{0.001}{0.051}$        &   $\underset{0.001}{0.096}$       &     $\underset{0.013}{10.120}$     &   $\underset{0.001}{0.197}$      \\[0.55em]
\textbf{CSE}  \cite{communitystructure} &$\underset{0.002}{0.229}$&$\underset{0.046}{\textbf{13.116}}$&$\underset{0.003}{0.070}$ & $\underset{0.001}{\textbf{0.026}}$&
$\underset{0.345}{20.314}$&
$\underset{\,\,\,0.006}{\textbf{-0.290}}$
 & $\underset{0.003}{\textbf{0.191}}$&
$\underset{0.055}{6.544}$&
$\underset{0.006}{\textbf{0.006}}$ &$\underset{0.002}{0.089}$
&$\underset{0.001}{6.554}$&
$\underset{0.001}{0.165}$    \\[0.55em]

\textbf{SP} \cite{shortest_path}  &$\underset{0.001}{0.221}$&$\underset{0.062}{12.842}$&$\underset{0.002}{0.106}$&$\underset{0.001}{0.037}$&$\underset{0.125}{23.451}$&$\underset{\,\,\,0.001}{-0.292}$&$\underset{0.001}{0.203}$&$\underset{0.032}{\textbf{7.258}}$&$\underset{0.002}{0.043}$&$\underset{0.001}{0.079}$&$\underset{0.007}{8.176}$&$\underset{0.001}{\textbf{0.209}}$   \\[0.55em]

\hline

\end{tabular}
}}

\end{table*}

\section{Conclusion and Future Directions}\label{sec:little_ball_of_fur_conclusion}
In this paper we described \textit{Little Ball of Fur} -- an open-source Python graph sampling framework built on the widely used scientific computing libraries NetworkX \cite{networkx}, NetworKit \cite{networkit}, and NumPy \cite{numpy}. In detail it provides techniques for node, edge, and exploration based graph sampling. 

We reviewed the general conventions which we used for implementing graph sampling algorithms in \textit{Little Ball of Fur}. The framework offers a limited number of public methods, ingests and outputs data in widely used graph formats, and embodies preset default hyperparameters. We presented practical implications of these desgin features with illustrative examples of Python code snippets. Using various social networks and web graphs we had shown that using sampled graphs extracted with \textit{Little Ball of Fur} one can approximate ground truth graph level statistics such as transitivity and the degree correlation coefficient. We also found evidence that sampling subgraphs with our framework can accelerate node and graph classification without extremely reducing the predictive performance.

As we have emphasized \textit{Little Ball of Fur} assumes that the inputted graph is undirected and unweighted. In the future we envision to relax these assumptions about the input. We plan to include additional high performance backend libraries such as SNAP \cite{snap} and GraphTool \cite{graphtool}. Furthermore, we aim to extend our framework by including multiplex \cite{gjoka2011multigraph}, attributed, and heterogeneous \cite{li2011sampling, yang2013semantically} graph sampling algorithms with new releases of the framework. 
\section*{Acknowledgements}
Benedek Rozemberczki was supported by the Centre for Doctoral Training in Data Science, funded by EPSRC (grant EP/L016427/1).
\bibliographystyle{ACM-Reference-Format}
\bibliography{main}

\end{document}